\documentclass[12pt,letterpaper]{article}
\usepackage{setspace}
\usepackage{color}
\usepackage{amsmath}
\usepackage{amsfonts}
\usepackage{verbatim}
\usepackage{amssymb}
\usepackage{graphicx,bm}
\usepackage{bbm}
\usepackage{cite}
\usepackage{authblk}
\usepackage{amsthm}
\usepackage{enumerate}
\usepackage{comment}
\usepackage{amssymb}
\usepackage{epstopdf}
\usepackage{epsf}
\usepackage{authblk}
\usepackage{tabularx, multirow, booktabs}
\usepackage{comment}
\usepackage{graphics}
\usepackage[dvipsnames,svgnames,x11names,table]{xcolor}
\usepackage[colorlinks = true,
            linkcolor = Blue,
            urlcolor  = Blue,
            citecolor = Blue,
            anchorcolor = Blue]{hyperref}

\date{}

\begin{document}

\title{\textbf{The solution space of a five-dimensional geometry: Kundt spacetimes and cosmological time-crystals}}
\author[1]{Homa Shababi\thanks{h.shababi@scu.edu.cn}}
\author[2]{Theophanes Grammenos\thanks{thgramme@uth.gr}}
\author[3]{Nikolaos Dimakis\thanks{nikolaos.dimakis@ufrontera.cl}}
\author[4]{Andronikos Paliathanasis\thanks{anpaliat@phys.uoa.gr}}
\author[5]{Theodosios Christodoulakis\thanks{tchris@phys.uoa.gr}}
\affil[1]{Center for Theoretical Physics, College of Physics, Sichuan University, Chengdu 610065, P. R. China}
\affil[2]{Department of Civil Engineering, University of Thessaly, 38334 Volos, Greece}
\affil[3]{Departamento de Ciencias F\'{\i}sicas, Universidad de la Frontera, Casilla 54-D, 4811186 Temuco, Chile}
\affil[4]{Institute of Systems Science, Durban University of Technology, Durban 4000, South Africa}
\affil[4]{Departamento de Matem\'{a}ticas, Universidad Cat\'{o}lica del Norte, Avda. Angamos 0610, Casilla 1280 Antofagasta, Chile}
\affil[4]{School for Data Science and Computational Thinking and Department of Mathematical Sciences, Stellenbosch University, Stellenbosch, 7602, South Africa}
\affil[5]{Nuclear and Particle Physics section, Physics Department, University of Athens, 15771 Athens, Greece}

\setcounter{Maxaffil}{0}
\renewcommand\Affilfont{\itshape\small}

\maketitle

\begin{abstract}
We uncover the solution space of a five dimensional geometry which we deem it as the direct counterpart of the Bianchi Type V cosmological model. We kinematically reduce the scale factor matrix and then, with an appropriate scaling and choice of time, we cast the spatial equations into a simple ``Kasner'' like form; thus revealing  linear integrals of motion. Their number is enough so that, along with the quadratic constraint,  it suffices to scan the entire space of solutions. The latter is revealed to be quite rich, containing cosmological solutions, some of which admit dimensional reduction asymptotically to four dimensions, Kundt spacetimes with vanishing type I (polynomial) curvature scalars and solutions describing periodic universes which behave like cosmological time crystals. 
\end{abstract}

\section{Introduction}

Although General Relativity addresses with success various gravitational
phenomena \cite{grt1,grt2,grt3}, it is challenged by the cosmological
observations \cite{rr1,Teg,Kowal,kkd1,kkd2}. The cosmological constant $%
\Lambda $ has been addressed to explain the current acceleration phase of
the universe. Nevertheless, the cosmological constant suffers from two
fundamental problems \cite{Peri08}, the \textquotedblleft old cosmological
constant problem\textquotedblright\ \cite{Weinberg89}, also known as
\textquotedblleft the fine tuning problem\textquotedblright , and
\textquotedblleft the coincidence problem\textquotedblright\ \cite%
{coincidence}. Nowadays the $H_{0}$-tension challenges further the nature of
the cosmological constant \cite{cos1}.

Brane cosmology has been used to provide mechanisms that explain
the aforementioned problems \cite{cc1,cc2,cc4}. In brane theory,
the physical spacetime  is assumed to be of higher-dimension, where the
observable four-dimensional universe is only a part of it \cite{br1,br2,br3,br4}. Isotropic and anisotropic brane-world cosmological models
have been considered in the literature in order to address the isotropy
problem and to describe the cosmological history \cite{bm0,bm1,bm2,bm3,bm4,bm5}.
In \cite{bm6} the asymptotic dynamics for brane-Mixmaster universe
was investigated, and it was found that the isotropic universe can be an
attractor for the cosmos's evolution. 

There are very few known exact solutions in the literature for anisotropic
brane cosmological models. Some exact solutions which describe brane-Bianchi
I geometries are determined in \cite{bbi1,bbi2}, while for some other
brane-Bianchi type exact solutions we refer the reader in \cite{bbi3,bbi4}. On
the other hand some exact solutions on branes geometries, which describe
astrophysical systems, can be found in \cite{ii1,ii2,ii3,ii4} and references
therein. There are also examples of higher dimensional spacetimes, where fundamental constants emerge when the spacetime reduces asymptotically to a four dimensional manifold \cite{Fukui}. The case of five-dimensional gravity is especially interesting due to Campbell's theorem, which states that any $n$-dimensional manifold can be locally embedded in $n+1$ dimensions. As a result, five-dimensional vacuum solutions can be realized as four-dimensional general relativistic solutions in the presence of matter \cite{Romero,Wesson1,Wesson2}.

In this study we are interested in the five dimensional equivalent of a Bianchi V geometry. In four dimensions, the Bianchi type V space of solutions in vacuum is completely known \cite{BV1,BV2,BV3}. Here, we consider a non-trivial extension in five dimensions, by considering a 4D algebra spanned by the spatial isometries, which has similar algebraic properties with the 3D Bianchi type V algebra of the four-dimensional cosmological model. We start from the most general metric, which - without loss of generality - we reduce to a simpler form. By adopting a particular time gauge we reveal the existence of certain integrals of motion, which we then use to integrate the vacuum equations. 

Among the solutions we obtain we distinguish spacetimes belonging to the Kundt class. The latter being Lorentzian manifolds that admit a null vector field that is geodesic, shear-free, twist-free, and expansion-free \cite{Kundt0,Kundt1}. Kundt spaces play a fundamental role in physical investigations as they include important geometric structures as subcases, like pp-waves and VSI (Vanishing Scalar Invariants) \cite{VSI} spacetimes which have found applications from pure gravitational physics to string theory \cite{Kpp1,Kpp2}. Especially, pp-waves form generalizations that include exact gravitational wave solutions, whose theoretical study has led to important results associated to phenomena like the memory effects in gravitational waves \cite{ppmem1,ppmem2,ppmem3,ppmem4,ppmem5,ppmem6}.

The structure of the paper is the following: In section \ref{sec2} we make a brief review of the canonical analysis in $d+1$ dimensions and write the general equations. In section \ref{sec3} we explain our choice of a 4D algebra to characterize the spatial isometries and make use of its automorphisms to simplify the form of the metric. Later, in section \ref{sec4} we write the resulting equations and we show that there exists a convenient time gauge which makes evident the existence of certain integrals of motion, which we use to solve the equations in their generality. We present the various solutions that we obtained and we explore their properties. Finally, section \ref{sec6} contains the sum of our results and a brief discussion of our findings.

\section{Canonical Analysis} \label{sec2}

We begin by considering the line element of a $(d+1)$-dimensional spacetime that admits a $d$-dimensional group of isometries acting simply transitively on the spacelike surfaces $t=$const.,
\begin{equation} \label{genlineel}
  ds^2 =  \left[- N^2(t) + N^\alpha(t) N_\alpha(t)\right] dt^2 + 2 N_\alpha(t) \sigma^\alpha_{\; i}(x) dt d x^i + \gamma_{\alpha\beta}(t) \sigma^\alpha_{\; i}(x) \sigma^\beta_{\; j}(x) dx^i dx^j,
\end{equation}
where the Latin $i,j$, and Greek $\alpha,\beta$, indices run from $1$ to $d$. The first are used to indicate the spatial coordinates, while the latter are internal indices related to the $d$-dimensional group of isometries. The $\sigma^\alpha=\sigma^\alpha_{\; i}(x)dx^i$ are the Cartan forms, having only spatial dependence on $x$, and satisfying
\begin{equation} \label{cforms}
  d \sigma^\alpha = - \frac{1}{2} C^{\alpha}_{\; \mu\nu} \sigma^\mu \wedge \sigma^\nu ,
\end{equation}
where the $C^{\alpha}_{\; \beta\mu}$ are the structure constants of the algebra whose elements are the generators of the isometry group. The temporal functions $N(t)$ and $N_\alpha(t)$ are the lapse function and shift vector respectively, while the $\gamma_{\alpha\beta}(t)$'s constitute the entries of the scale factor matrix.
The metric involved in the line element \eqref{genlineel} carries the decoupling of the variables $t$ and $x$ to the level of Einstein's field equations. Thus, the latter reduce to a system of coupled ordinary differential equations for the temporal functions $N$, $N_\alpha$ and $\gamma_{\alpha\beta}$:
\begin{subequations}  \label{eineq}
\begin{align}
  & { }^{{\scriptscriptstyle (d)}}\! R + K^2 - K_{\alpha\beta} K^{\alpha\beta} =0, \label{eineq1} \\
  & K^{\; \beta}_{\mu} C^{\mu}_{\alpha\beta} - K_{\alpha}^{\; \beta} C^{\mu}_{\;\beta \mu} = 0, \label{eineq2} \\
  & \dot{K}_{\alpha\beta} = N\; { }^{{\scriptscriptstyle (d)}}\! R_{\alpha\beta} - 2 N K_{\alpha \mu} K_{\beta}^{\; \mu} + N K K_{\alpha\beta} - N^{\mu} \left( K_{\alpha \lambda} C^{\lambda}_{\; \mu \beta} + K_{\lambda\beta} C^{\lambda}_{\; \mu \alpha} \right) \label{eineq3}.
\end{align}
\end{subequations}
In the latter relation, the dot signifies derivation with respect to the time variable $t$. The ${ }^{{\scriptscriptstyle (d)}}\! R$ is the Ricci scalar curvature of the $d$-dimensional spatial slices calculated by ${ }^{{\scriptscriptstyle (d)}}\! R = { }^{{\scriptscriptstyle (d)}}\! R_{\alpha\beta} \gamma^{\alpha\beta} $, where
\begin{equation}
  \begin{split}
    { }^{{\scriptscriptstyle (d)}}\! R_{\alpha\beta} = & - \frac{1}{2} C^{\kappa}_{\; \lambda\alpha} \left( C^{\lambda}_{\; \kappa\beta} + C^{\mu}_{\; \nu \beta} \gamma^{\lambda\nu} \gamma_{\kappa\mu} \right) +\frac{1}{4} \gamma^{\mu\nu} \gamma^{\rho\eta} C^{\kappa}_{\; \mu \rho} C^{\lambda}_{\; \nu \eta} \gamma_{\alpha\lambda} \gamma_{\kappa\beta} \\
    & + \frac{1}{2} C^{\mu}_{\; \mu\nu} \gamma^{\nu\rho} \left( C^{\kappa}_{\; \rho\alpha} \gamma_{\beta\kappa} + C^{\kappa}_{\; \rho_\beta} \gamma_{\alpha\kappa} \right)
  \end{split}
\end{equation}
is the corresponding Ricci tensor. Greek indices are raised and lowered with the use of the scale factor matrix, $\gamma_{\alpha\beta}$. The $K_{\alpha\beta}$ is the extrinsic curvature given by
\begin{equation} \label{extrcurv}
  K_{\alpha\beta}= - \frac{1}{2 N} \left( \dot{\gamma}_{\alpha\beta} + N^{\lambda} C^{\kappa}_{\; \lambda\beta} \gamma_{\alpha\kappa} + N^{\lambda} C^{\kappa}_{\; \lambda\alpha} \gamma_{\kappa\beta} \right), 
\end{equation}
and $K=K_{\alpha\beta} \gamma^{\alpha\beta}$. The equations \eqref{eineq1} and \eqref{eineq2} are the quadratic and linear constraints respectively, while \eqref{eineq3} is the set of the second order equations corresponding to the spatial part of Einstein's equations.

\section{Simplification of the metric} \label{sec3}

It is known that the symmetries of the spacetime can be used to reduce the nonzero components of the metric. Consider the following transformation
\begin{equation}\label{transf}
  \begin{split}
     t &\mapsto \tilde{t}=t, \\
     x^i &\mapsto \tilde{x}^i = g^i(t,x) \Rightarrow x^i = f^i(t,\tilde{x}).
  \end{split}
\end{equation}
In order to maintain the expression of a line element in which the homogeneity is manifest, we restrict the functions $f$ so that (see \cite{chris2001} for details )
\begin{subequations} \label{transfcond}
\begin{align}
   \sigma^\alpha_{\; i}(x) \frac{\partial x^i}{\partial \tilde{x}^m} & = \Lambda^\alpha_{\; \beta}(\tilde{t}) \sigma^\beta_{\; m}(\tilde{x}),\\
   \sigma^\alpha_{\; i}(x) \frac{\partial x^i}{\partial \tilde{t}} & = P^\alpha(\tilde{t}),
\end{align}
\end{subequations}
where we introduce the pure functions of time $\Lambda^\alpha_{\; \beta}(\tilde{t})$ and $P^\alpha(\tilde{t})$. With the above setting the line element \eqref{genlineel}, transforms to
\begin{equation}
  ds^2 =  \left[- \tilde{N}^2(\tilde{t}) + \tilde{N}^\alpha(\tilde{t}) \tilde{N}_\alpha(\tilde{t})\right] d\tilde{t}^2 + 2 \tilde{N}_\alpha(\tilde{t}) \sigma^\alpha_{\; i}(\tilde{x}) d\tilde{t} d \tilde{x}^i + \tilde{\gamma}_{\alpha\beta}(\tilde{t}) \sigma^\alpha_{\; i}(\tilde{x}) \sigma^\beta_{\; j}(\tilde{x}) d\tilde{x}^i d\tilde{x}^j,
\end{equation}
where
\begin{subequations} \label{taut}
\begin{align} \label{taut1}
   \tilde{\gamma}_{\alpha\beta} &= \Lambda^\mu_{\; \alpha} \Lambda^\nu_{\; \beta} \gamma_{\mu\nu}, \\
   \tilde{N}_\alpha  &= \Lambda^\mu_{\; \alpha} \left( N_\mu + P^\beta \gamma_{\beta\mu} \right), \\
   \tilde{N} & = N .
\end{align}
\end{subequations}
Such a transformation is possible if the following integrability conditions, emanating from \eqref{transfcond}, are satisfied:
\begin{subequations} \label{intcontr}
  \begin{align} \label{intcontr1}
    \Lambda^{\mu}_{\; \alpha} C^\alpha_{\; \kappa \lambda} & = \Lambda^{\alpha}_{\; \kappa} \Lambda^{\beta}_{\; \lambda} C^{\mu}_{\; \alpha\beta}, \\ \label{intcontr2}
    \Lambda^\mu_{\; \beta} C^{\alpha}_{\; \mu\nu} P^\nu & = \frac{d}{dt}\Lambda^{\alpha}_{\; \beta} .
  \end{align}
\end{subequations}
The general solution to the above two equations contains a number of constants and as many arbitrary functions of time as the dimension of the slice. These functions appear all in $P^\alpha$ and some or all in $\Lambda^\mu_\nu$. Thus, according to the occurrence, they can be used for simplifying $N_\alpha\,\gamma_{\mu\nu}.$
Equation \eqref{intcontr1} implies that the induced transformation is a time dependent automorphism of the algebra which characterizes the spatially homogeneous model under consideration. In many cases the optimal strategy to follow is to use the $\Lambda^\alpha_{\; \beta}$ in \eqref{taut1} to make zero as many components of the scale factor matrix as possible and then utilize the linear constraints to deduce the shift. In the subsequent subsection we make a brief description of the simplification process that we follow in the case of the $A^{1,1}_{4,5}$ algebra, according to the Patera classification \cite{Patera}.

\subsection{The $A^{1,1}_{4,5}$ algebra, automorphisms and reduction}

The reason behind the choice of this algebra of isometries for the spatial part of the metric is that it leads to a 4+1 generalization of the 3+1 Bianchi type V cosmological model while maintaining some key features: The automorphism matrix $\Lambda^{\alpha}_{\; \beta}$ is a direct generalization of the 3-dimensional type V case in what regards its form and the involved parameters \cite{tchris1}. Moreover, the maximum number of essential constants is the same as that of the $4A_1$ case (see \cite{tchris3} for its complete solution space) which forms the 5-dimensional generalization of the Bianchi type I model, in complete analogy to what happens in the 4-dimensional case \cite{tchris2}. Hence, we consider this algebra to produce the 4+1 generalization of the Bianchi type V cosmological model.

We consider as (initially) spatial coordinates the quadruplet $x^i=(x,y,z,w)$, where $i=1,\ldots,4$. The non-zero structure constants of the $A^{1,1}_{4,5}$ algebra are $C^{\alpha}_{\alpha4}=-C^{\alpha}_{4\alpha}=1$ in which, of course, $\alpha\neq 4$. The elements of the algebra, and differential generators of isometries for the spatial metric, are the four vectors
\begin{equation}\label{isometries}
  \xi_i = \frac{\partial}{\partial x^i},  \qquad  \xi_4 = x^i \frac{\partial}{\partial x^i} + \frac{\partial}{\partial w}, \qquad i=1,\ldots,3.
\end{equation}
The invariant one-forms are easily calculated to be
\begin{equation}\label{oneforms}
  \sigma^\alpha = e^w \delta^{\alpha}_{\; i} dx^i,  \qquad  \sigma^4 =dw, \qquad \alpha=1,\ldots,3.
\end{equation}
The automorphism matrix assumes the general form
\begin{equation}\label{automatrix}
  \Lambda^{\alpha}_{\; \beta} = \begin{pmatrix}
                                  a_1 & a_2 & a_3 & a_{10} \\
                                  a_4 & a_5 & a_6 & a_{11} \\
                                  a_7 & a_8 & a_9 & a_{12} \\
                                  0 & 0 & 0 & 1
                                \end{pmatrix},
\end{equation}
with the $a_I$ being arbitrary functions of $t$. Such a generic matrix satisfies the first set of integrability conditions, i.e. the equations given in \eqref{intcontr1}; in order for the second set  to be also satisfied we are led to the determination of the $P^\alpha$ functions and some of the $a_I$'s. In particular, for $I=2,\ldots,9$, we have $a_I(t) \propto a_1(t)$, while $a_1(t)$, $a_{10}(t)$, $a_{11}(t)$ and $a_{12}(t)$ remain arbitrary functions of time.The exact form of $P^\alpha$ is
\[
P^\alpha=\left\{\dot{a}_{10}(t)-\frac{a_{10}(t) \dot{a}_1(t)}{a_1(t)},\dot{a}_{11}(t)-\frac{a_{11}(t) \dot{a}_1(t)}{a_1(t)},\dot{a}_{12}(t)-\frac{a_{12}(t) \dot{a}_1(t)}{a_1(t)},-\frac{\dot{a}_1(t)}{a_1(t)}\right\}.
\]

The arbitrariness of $a_{10}(t)$, $a_{11}(t)$ and $a_{12}(t)$ can be used to eliminate the $\gamma_{14}$, $\gamma_{24}$ and $\gamma_{34}$ components of the scale factor matrix, while the arbitrariness of $a_1(t)$ can be utilized to set the determinant of its $3\times 3$ upper right block equal to the cubic power of $\gamma_{44}(t)$ . We thus have a setting
\begin{equation} \label{gamma}
 \gamma_{\mu\nu} = \begin{pmatrix}
                     \gamma_{11} & \gamma_{12} & \gamma_{13} & 0 \\
                     \gamma_{12} & \gamma_{22} & \gamma_{23} & 0 \\
                     \gamma_{13} & \gamma_{23} & \gamma_{33} & 0 \\
                     0 & 0 & 0 & \gamma_{44}
                   \end{pmatrix},
\end{equation}
where we choose to use the freedom of $a_1(t)$  in such a way so that $\det\left(\frac{\gamma_{\mu\nu}}{\gamma_{44}}\right)=1$, or equivalently, $\gamma_{44}$ being equal to the sub-determinant of the $3\times3$ block as seen in \eqref{gamma}, raised to the power of $1/3$, which leads to $\gamma=\det(\gamma_{\mu\nu})=\gamma_{44}^{4}$. In this setting the only freedoms left are the constant automorphisms (along with $P^\alpha=0$) that preserve the form of the scale factor matrix, i.e. the matrix $\Lambda^{\alpha}_{\; \beta}$ that we see in \eqref{automatrix} with $a_{10}=a_{11}=a_{12}=0$ and the rest of the $a_I$'s being constant. We now proceed to the next section where we determine the solutions to the vacuum field equations.

\section{The solution space} \label{sec4} 

\subsection{The integrals of motion}

By using the reduced scale factor matrix we indicated earlier, the structure constants of the algebra and of course, \eqref{extrcurv}, we can directly deduce from the linear constraint equations \eqref{eineq2} that the shift vector $N_\alpha$ has to be zero. This means that the relation for the extrinsic curvature simply reduces to $K_{\alpha\beta}=-\frac{1}{2N} \dot{\gamma}_{\alpha\beta}$. At the same time it can be easily checked that we have 
\[
 { }^{{\scriptscriptstyle (4)}}\! R_{\alpha\beta} = \frac{{ }^{{\scriptscriptstyle (4)}}\! R}{4} \gamma_{\alpha\beta},
 \]
 with ${ }^{{\scriptscriptstyle (4)}}\! R=-\frac{\gamma_{44}}{12}$.

With the previous consideration it can be easily seen that the spatial equations \eqref{eineq3} can now be written as
\begin{equation} \label{spared}
  \frac{d}{d t} \left( \gamma^{\rho\alpha} \dot{\gamma}_{\alpha\beta} \right) - \left( \frac{\dot{N}}{N}- \frac{\dot{\gamma}}{2 \gamma} \right) \gamma^{\rho\alpha} \dot{\gamma}_{\alpha\beta } + \frac{{ }^{{\scriptscriptstyle (4)}}\! R N^2}{2}  \delta^{\rho}_{\; \beta} =0,
\end{equation}
where we made use of Jacobi's formula $\gamma^{\kappa\lambda}\dot{\gamma}_{\kappa\lambda} = \dot{\gamma}/\gamma$. The second term of the above relation can be eliminated by simply choosing the time gauge $N= \sqrt{\gamma}$, where $\gamma$ is the determinant of the scale factor matrix. If then we take the ``44'' component of equation \eqref{spared}, due to $\gamma_{i4}=0$ for $i\neq 4$, we obtain
\begin{equation}
 \frac{d}{dt}\left(\frac{\dot{\gamma}_{44}}{\gamma_{44}} \right) + \frac{{ }^{{\scriptscriptstyle (4)}}\! R \gamma}{2}   =0.
\end{equation}
With this result we can finally write \eqref{spared} as
\begin{equation}\label{eqspa2}
  \frac{d}{d t} \left( \bar{\gamma}^{\rho\alpha} \dot{\bar{\gamma}}_{\alpha\beta} \right) =0,
\end{equation}
where
\begin{equation}
  \bar{\gamma}_{\alpha\beta} = \frac{1}{\gamma_{44}}\gamma_{\alpha\beta} ,
\end{equation}
is a scaled version of the scale factor matrix. We have thus demonstrated that in the gauge $N= \sqrt{\det(\gamma_{\mu\nu})}$ the second order equations possess integrals of motion of the form
\begin{equation}
   \bar{\gamma}^{\rho\alpha} \dot{\bar{\gamma}}_{\alpha\beta} = \theta^{\rho}_{\; \beta},
\end{equation}
with $\theta^{\rho}_{\; \beta}$ being a constant matrix whose form we are going to discuss right away. In matrix form we can write the above equation as $\bm{\bar{\gamma}}^{-1} \dot{\bm{\bar{\gamma}}}= \bm{\theta}$. 

Obviously, due to $\gamma_{\alpha\beta}$ (or equivalently $\bar{\gamma}_{\alpha\beta}$) being symmetric, we also need to demand, as a consistency requirement, that $\bm{\gamma \theta} = \bm{\theta}^{T} \bm{\gamma}$, where $\bm{\theta}^{T}$ is used for the transposed matrix. In addition to this, we may recall, from the previous section, that we still have the freedom of performing constant automorphisms, i.e. using the matrix \eqref{automatrix} where the first nine $a_I$ are constant entries and $a_{10}=a_{11}=a_{12}=0$ to induce transformations on the scale factor matrix $\bm{\gamma}$. These transformations are of the form $\bm{\gamma}\rightarrow \bm{\Lambda}^T \bm{\gamma \Lambda}$, which for the matrix $\bm{\theta}$ would imply $\bm{\theta}\rightarrow \bm{\Lambda}^{-1} \bm{\theta \Lambda}$, where $\bm{\Lambda}^{-1}$ is the inverse matrix of $\bm{\Lambda}$.  Due  to the relations $\text{det}(\bar{\gamma}_{\mu\nu})=1=\bar{\gamma}_{44}$ , the $\theta$ matrix is trace-less  $\theta^\mu_\mu=0$ with vanishing last line and column.

The entries of this $\boldsymbol{\theta}$ matrix are obviously real, therefore we have two main categories according to the nature of its eigenvalues being real or   imaginary. Further subdivisions follow for the cases of different, equal or zero eigenvalues. For the real category one can start by a triangular matrix
 \[
\boldsymbol{\theta} =\left(
\begin{array}{cccc}
\theta_{1}^{1} & \theta^{1}_{2} & \theta^{1}_{3} & 0 \\
 0 & \theta_{2}^{2} & \theta^{2}_{3} & 0 \\
 0 & 0 & -\theta_{1}^{1}-\theta_{2}^{2} & 0 \\
 0 & 0 & 0 & 0 \\
\end{array}
\right)
\]
and from it one can take the inequivalent irreducible forms presented in the following table (also containing the two cases of complex eigenvalues):

\begin{center}
\begin{tabularx}{\textwidth}{l@{\hskip 1in}l}
\toprule
Eigenvalues  & Form of the matrix\\
\toprule
Three non-zero different  & $\theta_1 =\left(
\begin{array}{cccc}
p_1 & 0 & 0 & 0 \\
 0 & p_2 & 0 & 0 \\
 0 & 0 & -p_1-p_2 & 0 \\
 0 & 0 & 0 & 0 \\
\end{array}
\right)$ \\
 \midrule
Three non-zero two equal   & $\theta_2=\left(
\begin{array}{cccc}
 p_2 & u & 0 & 0 \\
 0 & p_2 & 0 & 0 \\
 0 & 0 & -2 p_2 & 0 \\
 0 & 0 & 0 & 0 \\
\end{array}
\right)$\\
   \midrule
Three non-zero and equal    & $\theta_3 =\left(
\begin{array}{cccc}
 0 & 1 & 0 & 0 \\
 0 & 0 & 1 & 0 \\
 0 & 0 & 0 & 0 \\
 0 & 0 & 0 & 0 \\
\end{array}
\right)$\\
\midrule
Two non-zero different  &$\theta_4 =\left(
\begin{array}{cccc}
 p_1 & 0 & 0 & 0 \\
 0 & -p_1 & 0 & 0 \\
 0 & 0 & 0 & 0 \\
 0 & 0 & 0 & 0 \\
\end{array}
\right)$\\
\midrule
One non-zero real, two complex conjugate &$
\theta_5=\left(
\begin{array}{cccc}
 0 & 1 & 0 & 0 \\
 -p_1^2-p_2^2 & 2 p_1 & 0 & 0 \\
 0 & 0 & -2 p_1 & 0 \\
 0 & 0 & 0 & 0 \\
\end{array}
\right)$
\\
\midrule
One zero real, two imaginary conjugate&
$\theta_6=\left(
\begin{array}{cccc}
 0 & 1 & 0 & 0 \\
 -p_2^2 & 0 & 0 & 0 \\
 0 & 0 & 0 & 0 \\
 0 & 0 & 0 & 0 \\
\end{array}
\right)$
\\
        \bottomrule
        \end{tabularx}
\end{center}   


  
 \subsection{The Semi-Algorithm for solving the equations} 

After choosing a matrix $\theta$ from the table one must:

\begin{enumerate}

\item Solve the algebraic consistency relations $\bar{\gamma} \theta -\theta^T \bar{\gamma}=0$       \label{step1}

\item Solve the integrals of motion $\dot{\bar{\gamma}}=\bar{\gamma} \theta$ after using the acquired relations \label{step2}

\item Solve the Quadratic Constraint in terms of the only remaining unspecified scale factor, namely $\gamma_{44}(t)$ ($:=f(t)$ for the rest of the subsection).  \label{step3}
\end{enumerate}

Let us see in more detail the procedure by two examples for specific $\theta$'s:\\

\emph{Example 1}: $\theta_1$, three real, non-zero, different eigenvalues. \\

Step \ref{step1}, from the above list, dictates $\gamma_{12}(t)=\gamma_{13}(t)=\gamma_{23}(t)=0$, i.e. a diagonal scale factor matrix. If this information is inserted in step \ref{step2} we get  $\gamma_{11}(t)=f(t)\, e^{p_1 t}$, $\gamma_{22}(t)=f(t)\, e^{p_2 t}$,\\ $\gamma_{33}(t)=f(t)\,e^{-\left(p_1+p_2\right) t} $. The use of all these results in step \ref{step3} produces the following nonlinear differential equation of first order: $-6 \dot{f}(t)^2+\left(p_1^2+p_2 p_1+p_2^2\right) f(t)^2+24 f(t)^5=0$ its solution being $f(t)=\sqrt[3]{\frac{k}{24}} \text{cosh}^{-\frac{2}{3}}\left(\frac{1}{4} \sqrt{6 k} t\right)$ with $k=\left(p_1^2+p_2 p_1+p_2^2\right)$.\\

\emph{Example 2}: $\theta_3$, three real, equal eigenvalues (zero due to $\theta^\mu_\mu=0$).\\

Step \ref{step1} dictates $\gamma_{11}(t)=\gamma_{12}(t)=0\,\gamma_{22}(t)=\gamma_{13}(t)$. If this information is inserted in step \ref{step2} we get  
$\gamma_{13}(t)=f(t)$, $\gamma_{23}(t)=t\,f(t)$, $\gamma_{33}(t)=\frac{1}{2} t^2 f(t)$. The use of all these results in step \ref{step3} produces the following nonlinear differential equation of first order: $\dot{f}(t)^2-4 f(t)^5=0$ its solution being $f(t)=\pm \frac{1}{(3 t)^{2/3}}$. It is noteworthy that the corresponding geometry is a degenerate Kundt metric \cite{Kundt0,Kundt1} with an extra Killing and homothetic vector field. In all the above expressions the integration constants have been set to nominal values ($1$ or $0$) as they correspond to non-essential constants of the ensuing geometries.

\subsection{The Solutions}  

By following the previously described procedure we arrive to the following sets of solutions. We provide them in a form in which we have already absorbed any non-essential parameter in the line-element, with the use of diffeomorphisms.

At first, we obtain
\begin{equation} \label{sol1a}
  ds_{1}^2 = \frac{\mu}{\sinh^{\frac{2}{3}}t} \left[- \frac{dt^2}{9\sinh^2 t} + \sum_{i=1}^{3} e^{\kappa_i t-2 w} \left(d\mathrm{x}^i\right)^2 +dw^2  \right],
\end{equation}
where the three constants $\kappa_i$, $i=1,2,3$, are given by the set
\begin{equation}
  \kappa_i = \frac{2 \sqrt{\frac{2}{3}}}{\sqrt{p^2+p q+q^2}}  \left(p, q, -(p+q)\right), 
\end{equation}
while the $\mathrm{x}^i$ are the three spatial variables $(x,y,z)$. This metric is the case $d=4$ of a recently presented solution, see \cite{Sato}. Due to the spacetime symmetries, one can interchange the order of the spatial variables, e.g. $\mathrm{x}^i=(x,y,z)$ or $\mathrm{x}^i=(y,x,z)$, etc. This is also true for the subsequent solutions. It is interesting to note that under a complex transformation $t\rightarrow t + \mathrm{i} \frac{\pi}{2}$, with appropriate scalings in the spatial coordinates and $\mu$, we obtain the Euclidean counterpart of this solution which reads
\begin{equation}
  ds_{1(\text{Eucl.})}^2 = \frac{\mu}{\cosh^{\frac{2}{3}}t} \left[ \frac{dt^2}{9\cosh^2 t} + \sum_{i=1}^{3} e^{\kappa_i t-2 w} \left(d\mathrm{x}^i\right)^2 +dw^2  \right].
\end{equation}
These solutions emerge through the $\theta_1$ and for some special cases of the constants of integration, from $\theta_2$ and $\theta_4$.

\begin{figure}[ptb]
\includegraphics[width=\textwidth]{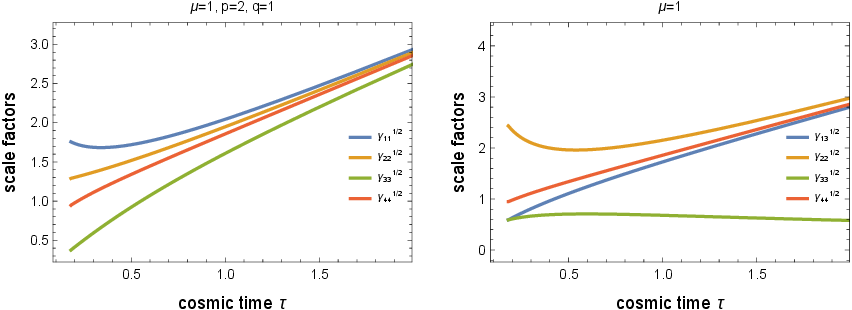}\caption{Evolution of the scale factors with respect to the cosmic time for the solutions \eqref{sol1a} and \eqref{sol2a} respectively.}%
\label{fig1}%
\end{figure}

The second line element we recover is
\begin{equation} \label{sol2a}
  ds_{2}^2 = \frac{\mu}{\sinh^{\frac{2}{3}}t} \left[- \frac{dt^2}{9 \sinh^2 t} + e^{-2 w} \left(e^{\pm\frac{4 \sqrt{2} }{3}t } dy^2 + e^{\mp\frac{2 \sqrt{2}}{3}  t} dz \left( dx + t dz \right)\right)+dw^2  \right],
\end{equation}
where once more one can interchange $(x,y,z)$ to alter the position of the nondiagonal term. As before, a complex time translation together with a more complicated complex transformation in the spatial variables, using the freedom of constant automorphisms, leads to the Euclidean line element that also forms a solution
\begin{equation}
  ds_{2(\text{Eucl.})}^2 = \frac{\mu}{\cosh^{\frac{2}{3}}t} \left[ \frac{dt^2}{9 \cosh^2 t} + e^{-2 w} \left(e^{\pm\frac{4 \sqrt{2} }{3}t } dy^2 + e^{\mp\frac{2 \sqrt{2}}{3}  t} dz \left(2 dx + t dz \right)\right)+dw^2  \right].
\end{equation}
These solutions correspond to cases involving the $\theta_1$ and $\theta_2$. Finally, from the $\theta_1$ we also obtain the flat space parametrized as
\begin{equation}
  ds_{3}^2 = \frac{1}{t^{\frac{2}{3}}} \left[ -\frac{ dt^2}{9 t^2} + e^{-2w} \left( dx^2 + dy^2 + dz^2\right) + dw^2  \right].
\end{equation}
The metrics that we read from \eqref{sol1a} and \eqref{sol2a} describe cosmological spacetimes and their signature is $(-,+,+,+,+)$; here and for the following line-elements we consider $\mu>0$. In both cases, if we calculate the cosmic time 
\begin{equation} \label{cosmictime}
\tau = \pm \int\!\! \sqrt{-g_{00}}d t, 
\end{equation}
we can see that the limit $t\rightarrow +\infty$ corresponds to the beginning of the universe $\tau \rightarrow 0$, while for $t\rightarrow 0$ we have $\tau \rightarrow +\infty$ \footnote{For the realization of this correspondence we need to choose the minus sign in \eqref{cosmictime} and also use the freedom of adding an arbitrary constant to an indefinite integral.}. There is an initial ``Big Bang'' singularity since the Kretschmann scalar $R^{\kappa\lambda\mu\nu}R_{\kappa\lambda\mu\nu}$ diverges in both cases when $t\rightarrow +\infty$ (i.e. $\tau \rightarrow 0$). In Fig. \ref{fig1} we demonstrate the evolution of the non-zero scale factors with respect to the cosmic time $\tau$. The graph on the left corresponds to the spacetime characterized by $ds^2_1$, for the values $\mu=1$, $p=2$, $q=1$; other values (positive or negative) also result to a similar expanding behaviour in all directions. On the other hand the graph on the right corresponds to $ds^2_2$ for $\mu=1$. We see that here there exists a contracting direction, the scale factor corresponding to $dz^2$. So, we may interpret this as a dimensional reduction \cite{dimred}, in the sense that we have the ``physical dimensions'' corresponding to $x,y,w$ expanding, while the extra $z$ direction contracts (although we do have an expansion in the $dx dz$ component, which we may however attribute it to the $dx$ part).

Continuing with the derivation of solutions, from $\theta_3$, we obtain the line elements
\begin{equation} \label{sol4a}
  ds_{4}^2 = \frac{1}{t^{\frac{2}{3}}} \left[-\frac{dt^2}{9 t^2} + e^{-2w} \left( dx^2+ 2 dz \left(t dx + dy \right) + \frac{t^2}{2} dz\right)+ dw^2 \right] , 
\end{equation}
with a homothetic vector
\begin{equation}
  \xi_{h} =t \frac{\partial}{\partial t}+y \frac{\partial}{\partial y}- z \frac{\partial}{\partial z},
\end{equation}
 and
\begin{equation} \label{sol5a}
  ds_{5} = \frac{1}{t^{\frac{2}{3}}} \left[ -\frac{dt^2}{9 t^2} + e^{-2w} \left(dx^2 + 2 dz \left( dx + dy \right) + t dz^2 \right) + dw^2\right].
\end{equation}
This metric admits the extra isometry and the homothecy
\begin{equation}
  \xi_{5} = z \frac{\partial}{\partial x} -\left(x+z\right) \frac{\partial}{\partial y}, \quad \xi_{h} =2 t \frac{\partial}{\partial t}+(x+y) \frac{\partial}{\partial y}- z \frac{\partial}{\partial z} .
\end{equation}
The last two geometries belong to the degenerate Kundt class \cite{Kundt0,Kundt1}, having zero all their polynomial scalar curvatures. This is seen by the existence of a null, geodesic and expansion-shear-twist free vector field $B=f(t,x,z,w)\,\frac{\partial}{\partial y}$ for both metrics. Interestingly enough, the last line element, $ds_5$, is a solution not only of the five dimensional equivalent of General Relativity, but also to the Einstein–Gauss–Bonnet gravity in five dimensions. The latter, apart from the Einstein-Hilbert part in the action, includes a contribution from the Gauss-Bonnet scalar $\mathcal{G}:=R^{\kappa\lambda\mu\nu}R_{\kappa\lambda\mu\nu}- 4 R^{\mu\nu}R_{\mu\nu} + R^2$ minimally coupled to the action. The $\mathcal{G}$ is a surface term in four dimensions, but contributes to the field equations from five dimensions and above. In our case, the family of metrics included in $ds_5$ (with interchanges in $x,y,z$) satisfies the Einstein-Gauss-Bonnet equations by making zero both $\mathcal{G}$ and its resulting contribution to the equations of motion, which is proportional to the tensor
\begin{equation}
  R R_{\mu\nu} -2 R_{\mu \kappa \nu \lambda} R^{\kappa\lambda} + R_{\mu \kappa\lambda\rho} R_{\nu}^{\; \kappa\lambda\rho} - 2 R_{\mu\kappa} R_{\nu}^{\;\kappa} - \frac{1}{4} g_{\mu\nu} \mathcal{G} .
\end{equation}
Solution $ds_{5}$ is the only one with this property; solving the field equations for both theories in vacuum. Cosmological solutions in Gauss-Bonnet theory have been extensively studied in the literature, mostly in what regards extensions of the FLRW model \cite{GB1,GB2,GB3,GB4}. The correspondence between $t$ and the cosmic time variable $\tau$ can be obtained in terms of an elementary function in the cases of \eqref{sol4a} and \eqref{sol5a}, $t\rightarrow \tau = t^{-1/3}$. With such a time transformation the line elements \eqref{sol4a} and \eqref{sol5a} become respectively
\begin{align}
  ds_4^2 &= - d\tau^2 + e^{-2w} \left( \tau^2 dx^2 + 2\left(\frac{1}{\tau} dx + \tau^2 dy\right) dz + \frac{1}{2 \tau^4} dz^2 \right) + \tau^2 dw^2 \\
  ds_5^2 &= - d\tau^2 + e^{-2w} \left( \tau^2 dx^2 + 2\tau^2 (dx+ dy) dz + \frac{1}{\tau}  dz^2 \right) + \tau^2 dw^2 .
\end{align}
We can see again that at infinity $\tau\rightarrow +\infty$, the $dz^2$ component vanishes, while all the other components in the metric expand to infinity. It is not obvious from the form of the line elements, but, by calculating the eigenvalues of the corresponding metrics we deduce that these spacetimes do not have a unique timelike direction, their signature contains an extra minus, i.e. $(-,-,+,+,+)$. By calculating the sub-determinant of the $2\times 2$ part of the $dy$ and $dz$ directions we see that it is negative, which implies that the extra minus in the signature is ``shared'' between these two directions.

From the family $\theta_5$ we arrive to the more complicated solution
\begin{equation} 
  \begin{split}
    ds_{6} = & \frac{\mu}{\cosh^{\frac{2}{3}}t} \Bigg\{ - \frac{dt^2}{9\cosh^2 t} +  e^{-2 w+ P t} \Bigg[  \Big( \left(p^2-q^2\right) \cos (Q\, t)+2 p q \sin (Q\, t)\Big) dx^2 \\
    & - 2 \Big(p \cos (Q\, t)+q \sin (Q \, t) \Big) dx dy + \cos (Q\, t) dy^2 + e^{-3 P t} dz^2\Bigg] +dw^2\Bigg\},
  \end{split}
\end{equation}
where the constants $P$, $Q$ are given in terms of $p$ and $q$ as 
\begin{equation}
  P = \frac{2 \sqrt{\frac{2}{3}} p}{\sqrt{3 p^2-q^2}}, \quad Q= \frac{2 \sqrt{\frac{2}{3}} q }{\sqrt{3 p^2-q^2}},
\end{equation}
with $3 p^2-q^2\neq0$. By using numerical values for $t$ and the involved parameters we can see that this is also a solution with signature  $(-,-,+,+,+)$, where however the $dt^2$ component is positive (assuming $\mu>0$), which means that we cannot consider this solution as a cosmological spacetime. The $t$ in this case rather plays the role of a spatial variable. The spacetime is singular for $t\rightarrow +\infty$.

\begin{figure}[ptb] \centering
\includegraphics{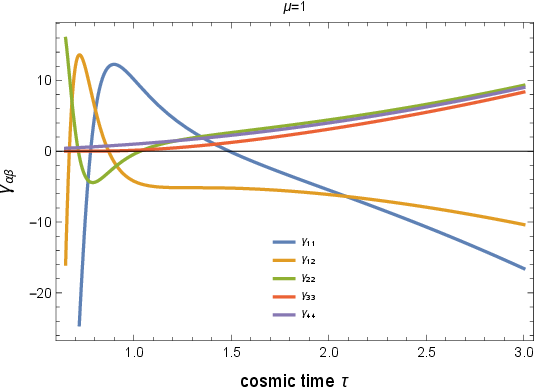}\caption{Evolution of the non-zero $\gamma_{\alpha\beta}$'s for the solutions \eqref{sol7b}.}%
\label{fig2}%
\end{figure}

When $q=\pm \sqrt{3} p$, instead of the previous solution, we are led to the line element
\begin{equation} \label{sol7a}
  \begin{split}
    ds_{7} =& \frac{\mu }{t^{2/3}} \Bigg[ -\frac{dt^2}{9 t^2} + e^{-2w} \Bigg( 2 e^{t}\left( \sqrt{3} \sin(\sqrt{3} t) - \cos(\sqrt{3} t)  \right) dx^2 
    \\ & - 2 e^{t}\left( \sqrt{3} \sin(\sqrt{3} t) + \cos(\sqrt{3} t)  \right) dxdy  + e^t \cos(\sqrt{3} t) dy^2\Bigg) +  dw^2 \Bigg].
  \end{split}
\end{equation} 
The transformation $t \rightarrow \tau= \sqrt{\mu} t^{-1/3}$ allows us to re-write it as
\begin{equation} \label{sol7b}
  \begin{split}
    ds_{7} =& -dt^2 + e^{-2w} \Bigg( 2 e^{\frac{\mu ^{3/2}}{\tau ^3}}\left( \sqrt{3} \sin(\sqrt{3} \frac{\mu ^{3/2}}{\tau ^3}) - \cos(\sqrt{3} \frac{\mu ^{3/2}}{\tau ^3})  \right) dx^2 
    \\ & - 2 e^{\frac{\mu ^{3/2}}{\tau ^3}}\left( \sqrt{3} \sin(\sqrt{3} \frac{\mu ^{3/2}}{\tau ^3}) + \cos(\sqrt{3} \frac{\mu ^{3/2}}{\tau ^3})  \right) dxdy  + e^{\frac{\mu ^{3/2}}{\tau ^3}} \cos(\sqrt{3} \frac{\mu ^{3/2}}{\tau ^3}) dy^2\Bigg)\\
    &  +  dw^2 \Bigg].
  \end{split}
\end{equation} 
The Kretschmann scalar is proportional to $\tau^{-13}$ which means that we encounter a curvature singularity at $\tau=0$. This metric is also of a $(-,-,+,+,+)$ signature (this time the extra minus appears in the $x-y$ subspace). The scale factors have a strong oscillatory behaviour at $\tau \rightarrow 0$ due to the involved trigonometric functions. In Fig. \ref{fig2} we present the curves of the various components of the spatial scale factor matrix with respect to the time variable $\tau$. They all have an expanding behaviour, even though some turn negative as $\tau\rightarrow +\infty$.

\begin{figure}[ptb] \centering
\includegraphics{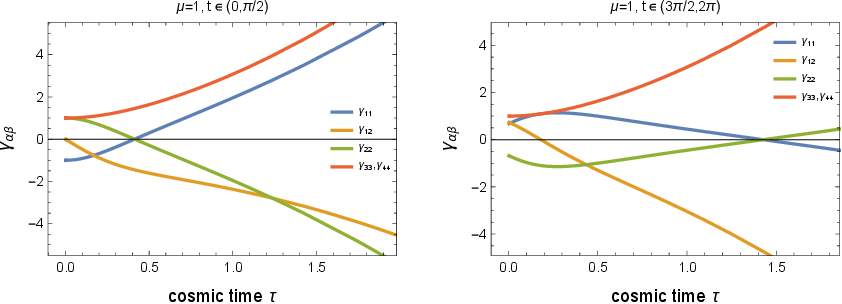}\caption{Evolution of the non-zero $\gamma_{\alpha\beta}$'s for the solutions \eqref{sol8} for two different regions of the time variable.}%
\label{fig3}%
\end{figure}

\begin{figure}[ptb] \centering
\includegraphics{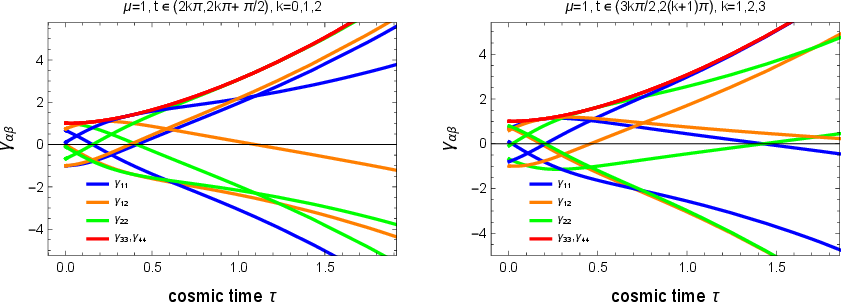}\caption{The change of $\gamma_{\alpha\beta}$'s for the solution \eqref{sol8} through different periods of the admissible regions of the time variable.}%
\label{fig4}%
\end{figure}

Finally, from the $\theta_6$ matrix we arrive at the solution
\begin{equation} \label{sol8}
  \begin{split}
    ds_8^2 = & \frac{\mu}{\cos^{\frac{2}{3}}t} \Bigg[\frac{- dt^2}{9\cos^2 t} - e^{-2w} \Bigg( \cos \left(2 \sqrt{\frac{2}{3}} t\right) \left( dx^2 - dy^2 \right) +2 \sin \left(2 \sqrt{\frac{2}{3}} t\right) dx dy \\
    & + dz^2\Bigg) + dw^2\Bigg]. 
  \end{split}
\end{equation}
This can also be recovered from $ds_{6}^2$ by setting $p=0$, $q=\mathrm{i}$ and scaling with the imaginary unit the $t$ and $x$ variables. By utilizing $\eqref{cosmictime}$ we can distinguish two disconnected regions in which the cosmic time runs in the real half line. The first is the region $t \in (0, \frac{\pi}{2})$ which maps to $\tau \in (0,+\infty)$. The second is the region $t \in (\frac{3\pi}{2},2 \pi)$, which also maps to $\tau \in (0,+\infty)$ with the difference that the limit $t\rightarrow \frac{3\pi}{2}$ corresponds to $\tau\rightarrow +\infty$ and $t\rightarrow 2\pi$ to $\tau\rightarrow 0$. What is more, $ds_8^2$ describes two periodic universes, each one with a period of $2\pi$, where however the $x-y$ components of the metric change in size after the completion of each $(2\pi k,2\pi k+\frac{\pi}{2})$, or $(2\pi k + \frac{3\pi}{2},2\pi (k+1))$, $k\in \mathbb{Z}$, period. In Fig. \ref{fig3} we observe the behaviour of the various scale factor matrix components for each such period. In Fig. \ref{fig4} we depict the pattern produced by the scale factors after three iterations. The solution \eqref{sol8}, as also the previous $ds_{7}^2$ case, is a spacetime of $(-,-,+,+,+)$ signature, but unlike $ds_{7}^2$, the Kretschmann scalar is not singular, so there is no apparent curvature singularity.  A similar pattern can be obtained in the regions  $t \in (\frac{\pi}{2},\pi)$ and  $t \in (\pi,\frac{3\pi}{2})$, but in order to do this we need to substitute in \eqref{sol8}, $(-\cos t)^{2/3}$ instead of $\cos^{2/3}t$. Again, each of the two regions (with the use of \eqref{cosmictime}) can be mapped to $\tau \in (0,+\infty)$ and describe an infinite number of universes with the addition of $2k \pi$. We refrain from presenting further these cases here since they are similar to the ones we already discussed.

Solutions of this type have been characterized in the literature as cosmological time crystals. The notion of a time crystal was introduced in \cite{time1,time2} for systems which exhibit periodic motion and the time translation symmetry is broken in their vacuum state (lowest energy state) \cite{time3}. In cosmology this characterization has been used for cyclic universes, mostly emerging in theories whose action includes higher order curvature terms \cite{time4,time5,time6}. Here, we obtain a vacuum solution describing an infinite number of spacetimes repeating cyclically, although the repetition is not entirely identical since in each cycle the starting values of the $x-y$ axes change. But we do have in effect a ``breaking'' of the time translation symmetry, not in the mathematical sense, since the mapping $t\rightarrow t+$const. still describes a solution, but in a physical one because this type of translation can send you from one distinct ``copy'' of a spacetime to another.

\section{Conclusion} \label{sec6}

We have investigated   and uncovered the space of solutions for a five dimensional model admitting the four-dimensional symmetry group corresponding  to the $A^{1\,1}_{4\,5}$ Lie algebra of  Patera's classification. The following facts permit us to attribute to the model the name ``Type V'': (a) only the trace of the structure constants tensor is non-zero ($c^i_{i4}=-c^i_{4i}=1,\,i=1,2,3$ no summation) ; (b) the ensuing automorphisms transformations matrix is the manifest generalization of the relevant matrix in the 4d Type V model both in shape and the number of parameters; (c) the counting of the maximum number for the essential parameters of the solutions 2x10 $\gamma_{\mu\nu}$'s -2 (for the Quadratic constraint)-4 (for the 4 linear constraints) -12 (free parameters of automorphisms)=2 is the same as that of the  $4 A_1$ (``Kasner'') model, in complete analogy to  what happens in the 4d cosmology (their common number being 1).

We have kinematically, through the time dependent automorphisms, reduced the scale factor matrix to a form such that the linear constraint equations dictate a zero shift ($N^\alpha=0$). At this stage an appropriate rescaling of the scale factor matrix ($\gamma_{\mu\nu} \rightarrow \frac{\gamma_{\mu\nu}}{\gamma_{44}}$) along with the usually very useful choice of time  $N=(\gamma)^{1/2}$ permits a complete first integration of the spatial equations by linear integrals of motion.Their number is such that all other $\gamma_{\mu\nu}$'s are  expressed in terms of $\gamma_{44}$. This turns the Quadratic constraint into a quadratic, first order differential equation for this last unknown function which is solved for all inequivalent (under the action of rigid automorphisms) irreducible forms of $\theta$.Thus, the entire solution space is revealed. 

We obtained solutions of various signatures. The majority of spacetimes can be characterized as cosmological since there is (at least one) timelike direction. In some cases we see that a dimensional reduction can be achieved asymptotically since you can have one contracting dimension, while the rest expand. We also distinguish the existence of Kundt spacetime solutions; the latter being generalizations of pp-waves, which have been investigated thoroughly in the context of higher order gravitational theories \cite{Kundt2,Kundt3}. In addition to the above we obtain a rather interesting time-periodic regular spacetime with two different admissible regions for the time variable, which alternate periodically and which we characterize as cosmological time crystals.

There is a prospect of application of the method we have developed here to other 5D or higher-dimensional models with appropriate symmetries.  We plan to employ geometric considerations to investigate higher-dimensional spatially homogeneous geometries which generalize their four-dimensional Bianchi counterparts. Such an analysis may reveal new directions in the study of higher-order theories, leading to intriguing geometric configurations like the ones developed here. Last but not least, the consideration of the Gauss-Bonnet term and Lovelock's theory in this context are of special interest. In a future work we intend to extend our research to these higher order generalizations of General Relativity.

\section*{Acknowledgements}

AP was supported by the UCN VRIDT through Resoluci\'on VRIDT No. $096/2022$ and Resoluci\'on VRIDT No. $098/2022$. AP thanks the support of National Research Foundation of South Africa.

\section*{Declarations}

\title{\textbf{Data Availability Statement}:} Data sharing is not applicable to this article as no datasets were generated or analyzed during the current study.\\
\title{\textbf{Funding and/or Conflicts of interests/Competing interests}:} The authors have no conflicts of interest to declare that are relevant to the content of this article.

\end{document}